\documentclass{svjour3}
\usepackage{amssymb}
\usepackage{amsmath}
\usepackage{amsfonts}

\smartqed  
\input{tcilatex}
\journalname{Foundations of Physics}

\begin{document}

%TCIMACRO{%
%\TeXButton{Title}{\title{Quantum Statistics of Identical Particles}}}%
%BeginExpansion
\title{Quantum Statistics of Identical Particles}%
%EndExpansion

%TCIMACRO{\TeXButton{Author(s)}{\author{J. C. Garrison}}}%
%BeginExpansion
\author{J. C. Garrison}%
%EndExpansion

%TCIMACRO{%
%\TeXButton{Institute}{\institute{J. C. Garrison \at
%Department of Physics, University of California at Berkeley  \email{jcg@berkeley.edu}
%}}}%
%BeginExpansion
\institute{J. C. Garrison \at
Department of Physics, University of California at Berkeley  \email{jcg@berkeley.edu}
}%
%EndExpansion

%TCIMACRO{%
%\TeXButton{Date}{\date{Received: 25 August, 2021/ Accepted: 27 May, 2022}}}%
%BeginExpansion
\date{Received: 25 August, 2021/ Accepted: 27 May, 2022}%
%EndExpansion

%TCIMACRO{\TeXButton{Make title}{\maketitle}}%
%BeginExpansion
\maketitle%
%EndExpansion

%TCIMACRO{\TeXButton{Begin abstract}{\begin{abstract}}}%
%BeginExpansion
\begin{abstract}%
%EndExpansion

The empirical rule that systems of identical particles always obey either
Bose or Fermi statistics is customarily imposed on the theory by adding it
to the axioms of nonrelativistic quantum mechanics, with the result that
other statistical behaviors are excluded \emph{a priori}. A more general
approach is to ask what other many-particle statistics are consistent with
the indistinguishability of identical particles. This strategy offers a way
to discuss possible violations of the Pauli Exclusion Principle, and it
leads to some interesting issues related to preparation of states and a
superselection rule arising from invariance under the permutation group.

%TCIMACRO{%
%\TeXButton{Keywords}{\keywords{many-body statistics \and nonrelativistic quantum theory \and Pauli exclusion principle \and quantum state preparation\and superselection rule }}}%
%BeginExpansion
%EndExpansion

%TCIMACRO{\TeXButton{PACS}{\PACS{ 03.65.-w \and 05.30.-d  }}}%
%BeginExpansion
%EndExpansion

%TCIMACRO{\TeXButton{MSC}{\subclass{20C35 \and 20B30\and 20C30 }}}%
%BeginExpansion
%EndExpansion

%TCIMACRO{\TeXButton{End abstract}{\end{abstract}}}%
%BeginExpansion
\end{abstract}%
%EndExpansion

\section{Introduction\label{&Intro}}

The strong empirical evidence that all systems of identical particles obey
either Bose or Fermi statistics is one of the most striking features of
quantum physics. An associated theoretical puzzle of long standing is that
these quantum-statistical properties of many-particle systems cannot be
derived from the axioms of quantum mechanics; instead, they must be imposed
by an additional assumption. In the seminal paper of Messiah and Greenberg 
\cite{MG} this is expressed as follows:

\begin{quote}
\textbf{Symmetrization Postulate\ (SP): }State vectors describing several
identical particles are either symmetric (bosons) or antisymmetric
(fermions) under permutations of the particle labels.\ 
\end{quote}

\noindent The familiar version of quantum mechanics--called SPQM below--is
defined by adding the Symmetrization Postulate to the axioms of
nonrelativistic quantum mechanics, \cite[Chap IIIb]{CT}. For $N$ identical
particles this means that the Hilbert space of states is either $\mathfrak{H}%
_{B}^{\left( N\right) }\ $(symmetric state vectors for bosons) or $\mathfrak{%
H}_{F}^{\left( N\right) }$ (antisymmetric state vectors for fermions).

The SPQM version of quantum theory has been phenomenally successful, but
even successful theories should be subjected to periodic experimental tests.
Any rule, \emph{e.g.} the Pauli Exclusion Principle (PEP)--that imposes
strict conditions on predictions of experimental results should be retested
as experimental techniques improve. This empirical motivation, combined with
the lack of any convincing theoretical explanation of the Symmetrization
Postulate, has stimulated a number of experimental searches for possible
SP-violations \cite{FSP1,FSP2,FSP3,FSP4,FSP5,FSP6,FSP7}. In view of the
continuing interest in these issues, a review of a more general theoretical
approach may lead to useful insights. In order to limit the theoretical
possibilities this generalization might allow, the following discussion will
be confined to a minimal extension of SPQM that combines the idea of
indistinguishability of identical particles with the axioms of quantum
mechanics \cite{MG}.

In what follows, a hermitian operator, $A$, that represents a measurable
physical quantity is called an \emph{observable}, and the set of its
eigenvalues is denoted by $ev\left( A\right) $. In general there can be
hermitian operators that do not represent any physical quantity; therefore,
a condition picking out those hermitian operators that may be observables
must be given for each physical system. An \emph{outcome}, $\left( a%
\boldsymbol{\ },\mathcal{E}_{a}\left( A\right) \right) $, of a measurement
of $A$ consists of the measured value, $a\in ev\left( A\right) $, and the
associated eigenspace, $\mathcal{E}_{a}\left( A\right) $, with the basis
set, 
\begin{equation}
\mathfrak{B}\left[ \mathcal{E}_{a}\left( A\right) \right] :=\left\{
\left\vert a:\mu \right\rangle \ |\ A\left\vert a:\mu \right\rangle
=a\left\vert a:\mu \right\rangle ,\ 1\leq \mu \leq d_{a}\ (\text{degeneracy
of }a)\ \right\}  \label{eigsp1}
\end{equation}%
For a set, $\boldsymbol{A}:=\left\{ A_{1},\ldots ,A_{J}\right\} $, of
compatible (commuting) observables the outcome, $\left( \boldsymbol{a\ },%
\mathcal{E}_{\boldsymbol{a}}\left( \boldsymbol{A}\right) \right) $, of a
joint measurement consists of the joint measured values, $\boldsymbol{a}%
=\left( a_{1},\ldots ,a_{J}\right) $, and the joint eigenspace,%
\begin{equation}
\mathcal{E}_{\boldsymbol{a}}\left( \boldsymbol{A}\right)
:=\dbigcap\limits_{j=1}^{J}\mathcal{E}_{a_{j}}\left( A_{j}\right) ,\dim 
\mathcal{E}_{\boldsymbol{a}}\left( \boldsymbol{A}\right) =d_{\boldsymbol{a}%
}\ \text{(degeneracy of }\boldsymbol{a}\text{)}  \label{joint-eigsp}
\end{equation}%
If the purpose of a measurement is to leave the system in a unique pure
state corresponding to a specific eigenvalue, it is called a preparation of
the state. Since this procedure involves rejecting outcomes with other
eigenvalues, it is also called filtering\emph{.}

In Section \ref{&Particles} the notion of identical particles is reviewed,
together with a description of systems of identical particles. A definition
of the indistinguishability of identical particles is presented in Section %
\ref{&Indis} and used to formulate an extension of SPQM that permits
violations of the Symmetrization Postulate. The special quantum features of
systems of indistinguishable particles are presented in Section \ref{&IP-QM}%
. Experiments seeking PEP-violations are briefly discussed in Section \ref%
{&Experiment}, together with a toy model that illustrates which features of
IPQM would be involved in their analysis. A general discussion is presented
in Section \ref{&IPQM}.

\section{Identical particles\label{&Particles}}

An `elementary particle' is currently understood to be one of the objects
listed in the Standard Model. Some bound states of the elementary particles,
e.g. nucleons, nuclei, atoms, and molecules, can also be regarded as
particles as long as they do not experience interactions strong enough to
excite their internal degrees of freedom. Each of these particles is
identified by a finite set of \emph{intrinsic\ properties} that are the
values of measurements on a single particle, e.g. mass, charge and spin.
This identification is assumed to be complete in the strong sense that there
are no additional single-particle measurements that could further define the
particle or distinguish between two particles with the same intrinsic
properties. Two particles are \emph{identical} if they have the same
intrinsic properties, and identical particles belong to the same \emph{%
species} \cite{IGK}. Each of these particles--with appropriate caveats for
quarks and gluons--can exist in the vacuum isolated from all other
particles. This property distinguishes them from those excited states of
many-particle systems that are called quasiparticles. Systems of
quasiparticles can display exotic forms of statistics \cite{Haldane,Leinaas}%
, but the present paper is intended to apply only to statistics arising from
the indistinguishability of identical particles.

The axioms of quantum theory do not provide any special rules for systems of
identical particles, but they do provide a framework for studying their
special properties. For this purpose, it is sufficient to consider a system
composed of only one particle species. The single-particle state space, $%
\mathfrak{H}_{spc}^{\left( 1\right) }$, is defined by a basis set, $%
\mathfrak{B}\left[ \mathfrak{H}_{spc}^{\left( 1\right) }\right] =\left\{
\left\vert \theta _{1}\right\rangle ,\left\vert \theta _{2}\right\rangle
,\ldots \right\} $, where each state identifier, $\theta ,$ is a set of
quantum numbers appropriate to the particle species in question, e.g. $%
\theta =\left( \boldsymbol{k},s\right) $ where $\boldsymbol{k}$ is a wave
number and $s\hslash $ is an eigenvalue of the $z$-component of the spin.
With the assignment of distinct \emph{particle labels}, $n=1,2,\ldots ,N$,
to the individual particles, the single-particle state space, $\mathfrak{H}%
_{n}^{\left( 1\right) }$, for particle $n$ is a copy of $\mathfrak{H}%
_{spc}^{\left( 1\right) }$, and single-particle state vectors in $\mathfrak{H%
}_{n}^{\left( 1\right) }$ are written as $\left\vert \psi \right\rangle _{n}$%
. Since no assumption about the behavior of state vectors under permutations
of the particle labels has been made, the $N$-particle state space, $%
\mathfrak{H}^{\left( N\right) }$, is the full tensor product of the
single-particle state spaces, $\mathfrak{H}^{\left( N\right) }=\mathfrak{H}%
_{1}^{\left( 1\right) }\otimes \cdots \otimes \mathfrak{H}_{N}^{\left(
1\right) }$, with the basis set,%
\begin{eqnarray}
\mathfrak{B}\left[ \mathfrak{H}^{\left( N\right) }\right] &=&\left\{
\left\vert \boldsymbol{\theta }\right\rangle :=\left\vert \theta _{1^{\prime
}}\right\rangle _{1}\otimes \cdots \otimes \left\vert \theta _{N^{\prime
}}\right\rangle _{N}\ |\ \boldsymbol{\theta }:=\left( \theta _{1^{\prime
}},\ldots ,\theta _{N^{\prime }}\right) \right\} ,  \notag \\
\left\langle \left. \boldsymbol{\kappa }\right\vert \boldsymbol{\theta }%
\right\rangle &:&=\left\langle \left. \kappa _{1^{\prime }}\right\vert
\theta _{1^{\prime }}\right\rangle _{1}\cdots \left\langle \left. \kappa
_{N^{\prime }}\right\vert \theta _{N^{\prime }}\right\rangle _{N}=\delta _{%
\boldsymbol{\kappa \theta }}:=\delta _{\kappa _{1^{\prime }}\theta
_{1^{\prime }}}\cdots \delta _{\kappa _{N^{\prime }}\theta _{N^{\prime }}}.
\label{BHN}
\end{eqnarray}%
The ordering convention is that the tensor products are always written in
increasing order of the particle labels. The primed subscripts $\left\{
1^{\prime },\ldots ,N^{\prime }\right\} $ in $\boldsymbol{\theta }$ are
called \emph{state labels}, since $n^{\prime }$ in $\left\vert \theta
_{n^{\prime }}\right\rangle _{n}$ identifies the single-particle state that
is assigned to particle $n$.

The $N!$ different assignments of labels to particles are related to each
other by the Symmetric Group, $\mathcal{S}_{N}$, composed of permutations, $%
n\rightarrow \mathcal{P}\left( n\right) $, of the labels $\left\{ 1,\ldots
,N\right\} $. For any pair of permutations $\mathcal{K}$ and $\mathcal{P}$
the product rule for the group $\mathcal{S}_{N}$ is $\left( \mathcal{KP}%
\right) \left( n\right) =\mathcal{K}\left( \mathcal{P}\left( n\right)
\right) $. Since the particle labels have no physical significance, a
permutation of the particle labels, with the state labels held fixed, is a
passive transformation (\emph{passive permutation}) analogous to a rotation
of the coordinate axes with the physical system held fixed. A passive
permutation, $\mathcal{P}$, acts on a basis vector $\left\vert \boldsymbol{%
\theta }\right\rangle $ in (\ref{BHN}) by first permuting the particle
labels--$\left\vert \theta _{n^{\prime }}\right\rangle _{n}\rightarrow
\left\vert \theta _{n^{\prime }}\right\rangle _{\mathcal{P}\left( n\right) }$%
--and then arranging the tensor product of the $\left\vert \theta
_{n^{\prime }}\right\rangle _{\mathcal{P}\left( n\right) }$'s in increasing
order of $\mathcal{P}\left( n\right) $. The combination of these two
operations is equivalent to leaving the particle labels unchanged while
performing the inverse permutation of the state labels:%
\begin{equation}
\mathcal{P}:\left\vert \theta _{1^{\prime }}\right\rangle _{1}\otimes \cdots
\otimes \left\vert \theta _{N^{\prime }}\right\rangle _{N}\rightarrow
\left\vert \theta _{\mathcal{P}^{-1}\left( 1^{\prime }\right) }\right\rangle
_{1}\otimes \cdots \otimes \left\vert \theta _{\mathcal{P}^{-1}\left(
N^{\prime }\right) }\right\rangle _{N}.  \label{PassivePerm}
\end{equation}%
A permutation of the state labels in $\boldsymbol{\theta }=\left( \theta
_{1^{\prime }},\ldots ,\theta _{N^{\prime }}\right) $, with the particle
labels held fixed, is an \emph{active permutation}, analogous to a rotation
of the physical system with the coordinate axes held fixed. Each active
permutation $\mathcal{P}$ acts on $\mathfrak{H}^{\left( N\right) }$ by an
operator, $D\left( \mathcal{P}\right) $, defined by: 
\begin{equation}
D\left( \mathcal{P}\right) \left\vert \boldsymbol{\theta }\right\rangle
:=\left\vert \widetilde{\mathcal{P}}\left( \boldsymbol{\theta }\right)
\right\rangle \text{ for all }\boldsymbol{\theta }\text{, where}\ \widetilde{%
\mathcal{P}}\left( \boldsymbol{\theta }\right) :=\left( \theta _{\mathcal{P}%
\left( 1^{\prime }\right) },\ldots ,\theta _{\mathcal{P}\left( N^{\prime
}\right) }\right) .  \label{PhatDef}
\end{equation}%
A straightforward argument shows that $D\left( \mathcal{P}\right) $ is a
unitary operator that satisfies, 
\begin{equation}
D\left( \mathcal{P}\right) D\left( \mathcal{K}\right) =D\left( \mathcal{PK}%
\right) ,
\end{equation}%
so that $\mathcal{P}\rightarrow D\left( \mathcal{P}\right) $ is a unitary
representation of $\mathcal{S}_{N}$ with carrier space $\mathfrak{H}^{\left(
N\right) }$. Comparing (\ref{PhatDef}) to (\ref{PassivePerm}) shows that the
effect of a passive permutation can be expressed as 
\begin{equation}
\mathcal{P}:\left\vert \boldsymbol{\theta }\right\rangle \rightarrow D\left( 
\mathcal{P}^{-1}\right) \left\vert \boldsymbol{\theta }\right\rangle ,
\label{PasAct}
\end{equation}%
i.e. the passive permutation $\mathcal{P}$ of the particle labels has the
same effect as the active permutation $\mathcal{P}^{-1}$ of the state labels.

\section{Indistinguishability of identical particles\label{&Indis}}

The assumption that there are no additional single-particle measurements
that can distinguish between two identical particles is the source of the
fundamental intuition that merely relabelling identical particles cannot
change any measurement outcomes. This means that two $N$-particle state
vectors related by a passive permutation of the particle labels must yield
the same probability distributions for the measured values of all
observables. Combining this with the relation (\ref{PasAct}) between passive
and active permutations suggests that the idea of the indistinguishability
of identical particles is captured by the following:

\begin{quote}
\textbf{Indistinguishability Postulate}: Two state vectors of several
identical particles that differ only by an active permutation of state
labels yield the same probability distributions for measurements of all
observables.
\end{quote}

\noindent This statement applies to all measurements on the entire system,
not to single-particle measurements on isolated particles. This is the
essential difference between indistinguishability and identity of particles.
The Indistinguishability Postulate and the Symmetrization Postulate are both
consistent with the axioms of quantum theory, but neither is derivable from
them. The version of quantum mechanics defined by adding the
Indistinguishability Postulate to the axioms will be called\textbf{\ }IPQM.
By virtue of its use of the word `observables,' the Indistinguishability
Postulate itself implies the rule that determines which hermitian operators
may be counted as observables.

For any state $\left\vert \Psi \right\rangle $ the probability of the
measurement outcome, $\left( a\boldsymbol{\ },\mathcal{E}_{a}\left( A\right)
\right) $, for $A$ is, 
\begin{equation}
prob\left( a|\Psi \right) =\sum_{\mu =1}^{d_{a}}\left\vert \left\langle
\left. a:\mu \right\vert \Psi \right\rangle \right\vert ^{2}=\sum_{\mu
=1}^{d_{a}}\left\langle \left. \Psi \right\vert a:\mu \right\rangle
\left\langle \left. a:\mu \right\vert \Psi \right\rangle =\left\langle \Psi
\left\vert \Pi \left( \mathcal{E}_{a}\left( A\right) \right) \right\vert
\Psi \right\rangle ,  \label{VN0}
\end{equation}%
where $\Pi \left( \mathcal{E}_{a}\left( A\right) \right) $ is the projection
operator onto $\mathcal{E}_{a}\left( A\right) $. For any observable $A$; any
eigenvalue $a\in ev\left( A\right) $; any pure state $\left\vert \Psi
\right\rangle $; and any active permutation; 
\begin{equation}
\left\vert \Psi \right\rangle \rightarrow \left\vert \Psi _{\mathcal{P}%
}\right\rangle =D\left( \mathcal{P}\right) \left\vert \Psi \right\rangle ;
\label{actP}
\end{equation}%
the Indistinguishability Postulate requires that the probability of finding $%
a$ in a measurement of $A$ is the same for the original and the permuted
state, 
\begin{equation}
prob\left( a|\Psi \right) =prob\left( a|\Psi _{\mathcal{P}}\right) .
\end{equation}%
Applying (\ref{VN0}) and (\ref{actP}) to both sides of this equation leads
to,%
\begin{equation}
\left\langle \Psi \left\vert \Pi \left( \mathcal{E}_{a}\left( A\right)
\right) \right\vert \Psi \right\rangle =\left\langle \Psi _{\mathcal{P}%
}\left\vert \Pi \left( \mathcal{E}_{a}\left( A\right) \right) \right\vert
\Psi _{\mathcal{P}}\right\rangle =\left\langle \Psi \left\vert D\left( 
\mathcal{P}\right) ^{\dagger }\Pi \left( \mathcal{E}_{a}\left( A\right)
\right) D\left( \mathcal{P}\right) \right\vert \Psi \right\rangle ,
\label{ProbK=ProbPerm}
\end{equation}%
for all $\left\vert \Psi \right\rangle \in \mathfrak{H}^{\left( N\right) }$
and all $\mathcal{P}\in \mathcal{S}_{N}$. Two operators with the same
expectation value in all states are equal; therefore, (\ref{ProbK=ProbPerm})
imposes a condition on the projection operator $\Pi \left( \mathcal{E}%
_{a}\left( A\right) \right) $ itself, 
\begin{equation}
\Pi \left( \mathcal{E}_{a}\left( A\right) \right) =D\left( \mathcal{P}%
\right) ^{\dagger }\Pi \left( \mathcal{E}_{a}\left( A\right) \right) D\left( 
\mathcal{P}\right) =\sum_{\mu =1}^{d_{a}}D\left( \mathcal{P}\right)
^{\dagger }\left\vert a:\mu \right\rangle \left\langle a:\mu \right\vert
D\left( \mathcal{P}\right) ,  \label{PiA=PiPdagAP}
\end{equation}%
for all $\mathcal{P}\in \mathcal{S}_{N}$. Since $D\left( \mathcal{P}\right) $
is unitary, the transformed operator, $A_{\mathcal{P}}:=D\left( \mathcal{P}%
\right) ^{\dagger }AD\left( \mathcal{P}\right) $, has the same eigenvalues
as $A$, and the corresponding transformed eigenvectors are $\left\vert 
\mathcal{P};a:\mu \right\rangle :=D\left( \mathcal{P}\right) ^{\dagger
}\left\vert a:\mu \right\rangle $. Substituting this into (\ref{PiA=PiPdagAP}%
) yields 
\begin{equation}
\Pi \left( \mathcal{E}_{a}\left( A\right) \right) =\sum_{\mu
=1}^{d_{a}}\left\vert \mathcal{P};a:\mu \right\rangle \left\langle \mathcal{P%
};a:\mu \right\vert =\Pi \left( \mathcal{E}_{a}\left( A_{\mathcal{P}}\right)
\right) .
\end{equation}%
The operators $A$ and $A_{\mathcal{P}}$ have the same eigenvalues and the
same eigenspaces; therefore, they are equal. Thus every observable $A$ must
satisfy, $A=A_{\mathcal{P}}=D\left( \mathcal{P}\right) ^{\dagger }AD\left( 
\mathcal{P}\right) $, which is equivalent to, 
\begin{equation}
\left[ A,D\left( \mathcal{P}\right) \right] =0,\text{for all }\mathcal{P}\in 
\mathcal{S}_{N}.  \label{PermInvOp}
\end{equation}%
An operator $A$ that satisfies this condition is said to be \emph{%
permutation-invariant}. Thus the Indistinguishability Postulate yields the,

\begin{quote}
\textbf{Permutation-Invariance Rule}: Observables for a system of identical
particles must be permutation-invariant.
\end{quote}

\noindent This is a necessary--but not sufficient--condition for a hermitian
operator to be an observable. An equivalent statement is: A hermitian
operator that does not satisfy (\ref{PermInvOp}) cannot be an observable. It
is important to remember that there can be hermitian operators that satisfy (%
\ref{PermInvOp}) but are not observables. These cases depend on particular
properties of the physical system under consideration. An important general
consequence of (\ref{PermInvOp}) is that for systems of identical particles
there are no observables that act on a single particle or any proper subset
of particles. Every observable must act on the entire system.

The behavior of state vectors under permutations is an equally important
feature of quantum theory for identical particles. A subspace $\mathfrak{N}$
of $\mathfrak{H}^{\left( N\right) }$ is called permutation-invariant\textbf{%
\ }if $D\left( \mathcal{P}\right) :\mathfrak{N}\rightarrow \mathfrak{N}$ for
all $\mathcal{P}\in \mathcal{S}_{N}$, or equivalently if $\Pi \left( 
\mathfrak{N}\right) $ is a permutation-invariant operator. Consequently, the
representation $\mathcal{P}\rightarrow D\left( \mathcal{P}\right) $ on $%
\mathfrak{H}^{\left( N\right) }$ induces a representation of $S_{N}$ on
every permutation-invariant subspace of $\mathfrak{H}^{\left( N\right) }$.
Since observables are permutation-invariant operators, each eigenspace of an
observable is a permutation-invariant subspace and thus a carrier space for
a representation of $S_{N}$. The expression (\ref{joint-eigsp}) shows that a
joint eigenspace of compatible observables is also a carrier space for a
representation of $\mathcal{S}_{N}$. A permutation-invariant subspace is
called irreducible if it has no proper permutation-invariant subspaces%
\footnote{%
A subspace is \emph{proper} if it is neither the whole space nor the null
subspace consisting of the zero vector alone}, and reducible otherwise. A
reducible subspace can be expressed as the direct sum of irreducible
subspaces \cite[Chap 3-13]{MH}. An irreducible (reducible) subspace carries
an irreducible (reducible) representation\textbf{\ }of $\mathcal{S}_{N}$ 
\cite[Chap. 7]{MH}. The irreducible representations of $\mathcal{S}_{N}$ are
denoted by $\mathcal{P}\rightarrow \mathfrak{D}^{\left( \gamma \right)
}\left( \mathcal{P}\right) $--or simply $\mathfrak{D}^{\left( \gamma \right)
}$--where $\mathfrak{D}^{\left( \gamma \right) }$ is a $d_{\gamma }\times $ $%
d_{\gamma }$ unitary matrix and $d_{\gamma }$ is the dimension of the
representation. The index $\gamma $ runs over the finite set $\Gamma _{N}$
defined in Appendix \ref{&IRepSN}.

\section{Quantum mechanics for indistinguishable particles\label{&IP-QM}}

In the SPQM version of quantum mechanics the state space, $\mathfrak{H}%
_{SP}^{\left( N\right) }$, for $N$ identical particles is either $\mathfrak{H%
}_{B}^{\left( N\right) }$ or $\mathfrak{H}_{F}^{\left( N\right) }$;
consequently, every $\left\vert \Psi \right\rangle \in \mathfrak{H}%
_{SP}^{\left( N\right) }$ satisfies $D\left( \mathcal{P}\right) \left\vert
\Psi \right\rangle =\sigma _{\mathcal{P}}\left\vert \Psi \right\rangle $,
where $\sigma _{\mathcal{P}}=1$ for bosons and $\sigma _{\mathcal{P}}=\left(
+1,-1\right) $ as $\mathcal{P}$ is (even, odd) for fermions. Since the
operators in quantum theory act on the Hilbert space of states--i.e. they
send the Hilbert space into itself--every operator $A$ acting on $\mathfrak{H%
}_{SP}^{\left( N\right) }$ satisfies, $D\left( \mathcal{P}\right)
A\left\vert \Psi \right\rangle =\sigma _{\mathcal{P}}A\left\vert \Psi
\right\rangle $, which in turn yields,%
\begin{equation}
\left[ D\left( \mathcal{P}\right) ,A\right] \left\vert \Psi \right\rangle
=D\left( \mathcal{P}\right) A\left\vert \Psi \right\rangle -AD\left( 
\mathcal{P}\right) \left\vert \Psi \right\rangle =0.  \label{PSauto}
\end{equation}%
Thus in SPQM all hermitian operators acting on $\mathfrak{H}_{SP}^{\left(
N\right) }$ automatically satisfy the permutation-invariance condition (\ref%
{PermInvOp}) required for observables.

By contrast, in the IPQM version of quantum mechanics the state space is the
complete tensor product $\mathfrak{H}^{\left( N\right) }$ of the
single-particle spaces; no symmetry conditions are imposed on the state
vectors. Instead, the permutation-invariance condition (\ref{PermInvOp}) is
used to pick out the hermitian operators that may be observables. Thus in
IPQM there are always hermitian operators acting on $\mathfrak{H}^{\left(
N\right) }$ that do not satisfy (\ref{PermInvOp}) and cannot be observables.

\subsection{Complete sets of compatible observables \label{&IP-Hspace}}

In quantum mechanics for distinguishable particles a \emph{complete set of
commuting operators}\ (CSCOP) is a finite set, $\boldsymbol{C}=\left\{
C_{1},\ldots ,C_{K}\right\} $, of mutually commutative hermitian operators
for which each joint eigenvalue is nondegenerate, so that each joint
eigenspace is one-dimensional \cite[Chap.IID3b]{CT}. In specific
applications a CSCOP is usually constructed from the relevant observables,
e.g. $\boldsymbol{C}=\left\{ H,L^{2},L_{z}\right\} $ for a scalar particle
moving in a central potential. Sets of hermitian operators satisfying this
definition can be constructed in many ways, but it is always implicitly
assumed that a CSCOP can be formed with observables. This is an important
issue for applications to state preparation.

In IPQM any attempt to apply the idea of a CSCOP to a system of
indistinguishable particles encounters the serious difficulty that there are
no CSCOPs. To see why, assume that a set, $\boldsymbol{A}$, of observables
is a CSCOP, then every eigenspace of $\boldsymbol{A}$ must be
one-dimensional. On the other hand, each eigenspace of $\boldsymbol{A}$ is a
carrier space for a representation of $\mathcal{S}_{N}$, and only the
symmetric and antisymmetric representations are one-dimensional, \emph{cf.}
Appendix \ref{&IRepSN}. Thus the unique basis vector, $\left\vert 
\boldsymbol{a}\right\rangle $, for each eigenspace would have to belong
either to $\mathfrak{H}_{B}^{\left( N\right) }$ or to $\mathfrak{H}%
_{F}^{\left( N\right) }$. The eigenstates defined by a CSCOP are supposed to
form a basis set for $\mathfrak{H}^{\left( N\right) }$; consequently, every\
vector $\left\vert \Psi \right\rangle $ in $\mathfrak{H}^{\left( N\right) }$
would be of the form, 
\begin{equation}
\left\vert \Psi \right\rangle =\left\vert \Psi :B\right\rangle +\left\vert
\Psi :F\right\rangle ,  \label{PsiBF}
\end{equation}%
where $\left\vert \Psi :B\right\rangle \in \mathfrak{H}_{B}^{\left( N\right)
}$ and $\left\vert \Psi :F\right\rangle \in \mathfrak{H}_{F}^{\left(
N\right) }$. This is true for $N=2$, but false for $N\geq 3$. If $\mathcal{P}
$ is an even permutation, then $D\left( \mathcal{P}\right) $ would leave all
vectors satisfying (\ref{PsiBF}) invariant; consequently, any $\left\vert
\Psi \right\rangle \in \mathfrak{H}^{\left( N\right) }$ that is not
invariant under even permutations is a counter example to (\ref{PsiBF}). For 
$N\geq 3$ suppose the components of $\boldsymbol{\theta }=\left( \theta
_{1^{\prime }},\ldots ,\theta _{N^{\prime }}\right) $ are all distinct, then
the basis vector $\left\vert \boldsymbol{\theta }\right\rangle $ is not
invariant under even permutations, or indeed any permutations at all;
therefore, for $N\geq 3$ there are no CSCOPs for systems of
indistinguishable particles described by IPQM. On the other hand, this
argument does not apply to SPQM, since all state vectors are in either $%
\mathfrak{H}_{B}^{\left( N\right) }$or $\mathfrak{H}_{F}^{\left( N\right) }$
to begin with. Thus CSCOPs are not forbidden in SPQM.

The physical necessity of preparing states by measurements of observables
means that any replacement for the CSCOP idea will still involve joint
measurements of some finite set, $\boldsymbol{A}=\left\{ A_{1},\ldots
,A_{K}\right\} $, of compatible observables. The definition of the joint
eigenspace, $\mathcal{E}_{\boldsymbol{a}}\left( \boldsymbol{A}\right) $, in (%
\ref{joint-eigsp}) guarantees that 
\begin{equation}
\dim \left[ \mathcal{E}_{\boldsymbol{a}}\left( \boldsymbol{A}\right) \right]
\leq \dim \left[ \mathcal{E}_{a_{k}}\left( A_{k}\right) \right] \ \
k=1,\ldots ,K.
\end{equation}%
With this in mind, it is useful to restrict the observables in $\boldsymbol{A%
}$ by requiring that at least one of them, say $A_{1}$, has only finite
degeneracies, i.e. every eigenspace of $A_{1}$ is finite-dimensional. When
this requirement is satisfied, the joint eigenspaces $\mathcal{E}_{%
\boldsymbol{a}}\left( \boldsymbol{A}\right) $ are all finite dimensional
carrier spaces for reducible or irreducible representations of $\mathcal{S}%
_{N}$.

Irreducibility is automatic for nondegenerate eigenvalues, since the
eigenspaces are one-dimensional. If $\boldsymbol{a}$ is degenerate and $%
\mathcal{E}_{\boldsymbol{a}}\left( \boldsymbol{A}\right) $ is irreducible, a
natural question is whether the $d_{\boldsymbol{a}}$-fold degeneracy can be
reduced by enlarging $\boldsymbol{A}$ to $\left\{ A_{1},\ldots
,A_{K},Z\right\} $, where $Z$ is a permutation-invariant hermitian operator
that commutes with all members of $\boldsymbol{A}$. Since $\mathcal{E}_{%
\boldsymbol{a}}\left( \boldsymbol{A}\right) $\ is invariant under $Z$, there
will be permutation-invariant eigenspaces, $\mathcal{E}_{z}\left( Z\right) $%
, that are subspaces of $\mathcal{E}_{\boldsymbol{a}}\left( \boldsymbol{A}%
\right) $, but the irreducible space $\mathcal{E}_{\boldsymbol{a}}\left( 
\boldsymbol{A}\right) $ does not contain any proper, permutation-invariant
subspaces. This leaves two possibilities: either $\mathcal{E}_{z}\left(
Z\right) =\left\{ \boldsymbol{0}\right\} $, or $\mathcal{E}_{z}\left(
Z\right) =\mathcal{E}_{\boldsymbol{a}}\left( \boldsymbol{A}\right) $. The
first is trivial and the second means that every vector in $\mathcal{E}_{%
\boldsymbol{a}}\left( \boldsymbol{A}\right) $ is an eigenvector of $Z$ with
a common eigenvalue $z$. Thus for any permutation-invariant, hermitian
operator $Z$ that is compatible with $\boldsymbol{A}$, 
\begin{equation}
\left\langle \boldsymbol{a}:\mu \left\vert Z\right\vert \boldsymbol{a}:\mu
^{\prime }\right\rangle =z\ \delta _{\mu \mu ^{\prime }}\ \left( \mu ,\mu
^{\prime }=1,\ldots ,d_{\boldsymbol{a}}\right) .  \label{ProIden}
\end{equation}%
Therefore, no measurement of an observable compatible with all the
observables in $\boldsymbol{A}=\left\{ A_{1},\ldots ,A_{K}\right\} $ can
distinguish between the pure states in an irreducible eigenspace of $%
\boldsymbol{A}$. This is the physical significance of irreducibility with
respect to the Symmetric Group.

If $\boldsymbol{a}$ is degenerate and $\mathcal{E}_{\boldsymbol{a}}\left( 
\boldsymbol{A}\right) $ is reducible, the $d_{\boldsymbol{a}}$-fold
degeneracy can be partially resolved by expressing $\mathcal{E}_{\boldsymbol{%
a}}\left( \boldsymbol{A}\right) $ as the direct sum of irreducible subspaces,%
\begin{equation}
\mathcal{E}_{\boldsymbol{a}}\left( \boldsymbol{A}\right)
=\dbigoplus\limits_{i=1}^{n_{\boldsymbol{a}}}\mathcal{E}_{\boldsymbol{a}%
}^{\left( i\right) }\left( \boldsymbol{A}\right) ,\ \ \ d_{\boldsymbol{a}%
}=\sum_{i=1}^{n_{\boldsymbol{a}}}d_{\boldsymbol{a}i},  \label{RedE}
\end{equation}%
where $\mathcal{E}_{\boldsymbol{a}}^{\left( i\right) }\left( \boldsymbol{A}%
\right) $ is a carrier space for an an irreducible representation $\mathfrak{%
D}^{\left( \gamma _{i}\right) }$. According to (\ref{ProIden}) it is not
possible to resolve the irreducible eigenspaces $\mathcal{E}_{\boldsymbol{a}%
}^{\left( i\right) }\left( \boldsymbol{A}\right) $ by measuring any
compatible observable, but it is possible to label them by using the
permutation-invariant, hermitian operator, 
\begin{equation}
Z:=\sum_{\boldsymbol{a}\in ev\left( \boldsymbol{A}\right) }\sum_{i=1}^{n_{%
\boldsymbol{a}}}z_{\boldsymbol{a}i}\Pi \left( \mathcal{E}_{\boldsymbol{a}%
}^{\left( i\right) }\left( \boldsymbol{A}\right) \right) ,
\end{equation}%
where the $z_{\boldsymbol{a}i}$'s are real and distinct. Every vector $%
\left\vert \psi \right\rangle $ in each irreducible subspace $\mathcal{E}_{%
\boldsymbol{a}}^{\left( i\right) }\left( \boldsymbol{A}\right) $ satisfies $%
Z\left\vert \psi \right\rangle =z_{\boldsymbol{a}i}\left\vert \psi
\right\rangle $; therefore, every eigenspace of the extended set, $\left\{
A_{1},\ldots ,A_{K},Z\right\} $, is irreducible.

This argument establishes the existence of a permutation-invariant,
hermitian operator that ensures irreducibility for every eigenspace of the
extended set, but there is no guarantee that this operator represents a
measurable quantity. Just as in the case of CSCOPs, the existence of an
observable that has the same effect as $Z$ must be assumed. When this is
true, the extended set is an example of a \emph{Complete Set of Compatible
Observables }(CSCOB), which is defined as a collection, $\boldsymbol{A}%
=\left\{ A_{1},\ldots ,A_{J}\right\} $, of compatible observables for which
the eigenspace in each outcome, $\left( \boldsymbol{a},\mathcal{E}_{%
\boldsymbol{a}}\left( \boldsymbol{A}\right) \right) $, of a joint
measurement of $\boldsymbol{A}$ carries an irreducible representation $%
\mathfrak{D}^{\left( \gamma _{\boldsymbol{a}}\right) }$ of $\mathcal{S}_{N}$%
. The irreducible eigenspaces obtained by measuring a CSCOB play the role of
the pure states obtained by measurement of a CSCOP for distinguishable
particles.

The eigenvalue set $ev\left( \boldsymbol{A}\right) $ for a CSCOB naturally
decomposes into subsets labelled by $\gamma \in \Gamma _{N}$, 
\begin{equation}
ev\left( \boldsymbol{A},\gamma \right) :=\left\{ \boldsymbol{a}\in ev\left( 
\boldsymbol{A}\right) \ |\ \gamma _{\boldsymbol{a}}=\gamma \right\} ,
\end{equation}%
i.e. for $\boldsymbol{a}\in ev\left( \boldsymbol{A},\gamma \right) $ the
eigenspace $\mathcal{E}_{\boldsymbol{a}}\left( \boldsymbol{A}\right) $ is a
carrier space for the irreducible representation $\mathfrak{D}^{\left(
\gamma \right) }$, with the basis set%
\begin{equation}
\mathfrak{B}\left[ \mathcal{E}_{\boldsymbol{a}}\left( \boldsymbol{A}\right) %
\right] =\left\{ \left\vert \gamma ,\boldsymbol{a}:g\right\rangle \ |\
g=1,\ldots ,d_{\gamma }\ \right\} ,  \label{BEagam}
\end{equation}%
where the basis vectors transform by,%
\begin{equation}
D\left( \mathcal{P}\right) \left\vert \gamma ,\boldsymbol{a}:g\right\rangle
=\sum_{g^{\prime }=1}^{d_{\gamma }}\left\vert \gamma ,\boldsymbol{a}%
:g^{\prime }\right\rangle \mathfrak{D}_{g^{\prime }g}^{\left( \gamma \right)
}\left( \mathcal{P}\right) .  \label{Ponvecga}
\end{equation}%
A superposition (mixture) of basis vectors $\left\vert \gamma ,\boldsymbol{a}%
:g\right\rangle $ with a common value of $\gamma $ is said to be a pure
(mixed) state of \emph{symmetry type}\textbf{\ }$\gamma $. Superpositions
(mixtures) of states of several distinct symmetry types are called \emph{%
hybrid-symmetry} pure (mixed) states. Every $\left\vert \Psi \right\rangle
\in \mathfrak{H}^{\left( N\right) }$ can be expressed as a superposition of
states of different symmetry types, 
\begin{equation}
\left\vert \Psi \right\rangle =\sum_{\gamma \in \Gamma _{N}}C_{\gamma
}\left\vert \Psi ^{\gamma }\right\rangle ;\ \text{where }\left\vert \Psi
^{\gamma }\right\rangle :=\sum_{\boldsymbol{a}\in ev\left( \boldsymbol{A}%
,\gamma \right) }\sum_{g=1}^{d_{\gamma }}\Psi ^{\gamma \boldsymbol{a}%
g}\left\vert \gamma ,\boldsymbol{a}:g\right\rangle ,  \label{expgam88}
\end{equation}%
\begin{equation}
\left\langle \left. \Psi ^{\gamma }\right\vert \Psi ^{\gamma }\right\rangle
=\sum_{\boldsymbol{a}g}\left\vert \Psi ^{\gamma \boldsymbol{a}g}\right\vert
^{2}=1,\ \text{and}\ \left\langle \left. \Psi \right\vert \Psi \right\rangle
=\sum_{\gamma }\left\vert C_{\gamma }\right\vert ^{2}=1.  \label{norm33}
\end{equation}

\subsection{State preparation\label{&Filtering}}

In SPQM a pure state $\left\vert \Psi \right\rangle $ that is a unique
eigenvector of a set of compatible observables is said to be a \emph{%
preparable pure state}. Preparation of states is, therefore, an important
application of CSCOPs that are composed of observables. If a physical system
has such a CSCOP and $\left\vert \Psi \right\rangle $ is one of unique
eigenvectors of $\boldsymbol{C}$, then $\left\vert \Psi \right\rangle $ is
prepared by measuring $\boldsymbol{C}$ and accepting the outcome with the
joint eigenvalue corresponding to $\left\vert \Psi \right\rangle $. Since
systems of indistinguishable particles under IPQM do not support CSCOPs,
this description of state preparation has to be worked out anew.

A measurement of a set, $\boldsymbol{A}$, of compatible observables leaves
the system in one of the eigenspaces of $\boldsymbol{A}$. If this eigenspace
happens to be one-dimensional, then the unique basis vector $\left\vert 
\boldsymbol{a}\right\rangle $ is a preparable pure state. For
indistinguishable particles, the one-dimensional eigenspaces of $\boldsymbol{%
A}$ carry the symmetric or antisymmetric representation of $\mathcal{S}_{N}$%
; therefore, all preparable pure states are either in $\mathfrak{H}%
_{B}^{\left( N\right) }$ or in $\mathfrak{H}_{F}^{\left( N\right) }$. This
severe restriction of the set of preparable pure states suggests that in
IPQM the idea of state preparation should be extended to mixed states
described by density operators. Since observables are permutation-invariant,
every observable $Y$ satisfies, 
\begin{equation}
Tr\left[ \varrho Y\right] =Tr\left[ \varrho D^{\dagger }\left( \mathcal{P}%
\right) YD\left( \mathcal{P}\right) \right] =Tr\left[ D\left( \mathcal{P}%
\right) \varrho D^{\dagger }\left( \mathcal{P}\right) Y\right] ,
\end{equation}%
for any density operator $\varrho $ and all $\mathcal{P}\in \mathcal{S}_{N}$
. Summing both sides of this equation over $\mathcal{P}\in \mathcal{S}_{N}$
yields $Tr\left[ \varrho Y\right] =Tr\left[ \bar{\rho}Y\right] $, where,%
\begin{equation}
\text{ }\bar{\rho}:=\frac{1}{N!}\sum_{\mathcal{P}\in \mathcal{S}_{N}}D\left( 
\mathcal{P}\right) \rho D^{\dagger }\left( \mathcal{P}\right) ,
\label{DenOpPI}
\end{equation}%
is a permutation-invariant density operator. Thus no generality is lost by
requiring physical density operators to be permutation-invariant.

Let $\boldsymbol{A}$ be a CSCOB, then a measurement with outcome $\left( 
\boldsymbol{a},\mathcal{E}_{\boldsymbol{a}}\left( \boldsymbol{A}\right)
\right) $ leaves the system in a state described by a density operator of
the form,%
\begin{equation}
\rho =\sum_{g=1}^{d_{\gamma }}\sum_{g^{\prime }=1}^{d_{\gamma }}\left\vert
\gamma ,\boldsymbol{a}:g\right\rangle \rho _{gg^{\prime }}\left\langle
\gamma ,\boldsymbol{a}:g^{\prime }\right\vert .  \label{rhoagam}
\end{equation}
Since the basis vectors $\left\vert \gamma ,\boldsymbol{a}:g\right\rangle $
are joint eigenvectors of $\boldsymbol{A}$, the operator $\rho $ commutes
with $\boldsymbol{A}$; consequently, (\ref{ProIden}) implies that $\rho
_{gg^{\prime }}\propto \delta _{gg^{\prime }}$. Combining this with the unit
trace condition yields, 
\begin{equation}
\varrho =\frac{1}{d_{\gamma }}\sum_{g=1}^{d_{\gamma }}\left\vert \gamma ,%
\boldsymbol{a}:g\right\rangle \left\langle \gamma ,\boldsymbol{a}%
:g\right\vert =\frac{1}{d_{\gamma }}\Pi \left( \mathcal{E}_{\boldsymbol{a}%
}\left( \boldsymbol{A}\right) \right) .  \label{maxprep}
\end{equation}%
Thus a measurement of the CSCOB\ $\boldsymbol{A}$ leaves the system in a
state described by a unique density operator. This state is called a \emph{%
preparable mixed state}. The result (\ref{maxprep}) is also called\ a maximal%
\emph{\ }state preparation or a maximal\emph{\ }filtering\textbf{\ }\cite{MG}%
. This is the best that can be done for indistinguishable particles
described by IPQM.

\subsection{Superselection rule\label{&SSR}}

The combination of the permutation-invariance rule with the first two of
Schur's Lemmas, \emph{cf.} Appendix \ref{&SchurL}, is the basis for a proof
of the following result: \emph{A permutation-invariant operator, }$X$\emph{,
cannot connect states of different symmetry types }\cite{MG}\emph{. }%
Consequently, for any permutation-invariant operator $X$, 
\begin{equation}
\left\langle \gamma ^{\prime },\boldsymbol{a}^{\prime }:g^{\prime
}\left\vert X\right\vert \gamma ,\boldsymbol{a}:g\right\rangle =\delta
_{\gamma ^{\prime }\gamma }\ M_{g^{\prime }g}^{\gamma }\left( X\right) ,
\label{nocon}
\end{equation}%
where,%
\begin{equation}
M_{g^{\prime }g}^{\gamma }\left( X\right) :=\left\langle \gamma ,\boldsymbol{%
a}^{\prime }:g^{\prime }\left\vert X\right\vert \gamma ,\boldsymbol{a}%
:g\right\rangle ;\ \left( g^{\prime },\ g=1,\ldots ,d_{\gamma }\right) .
\label{Xid}
\end{equation}%
The permutation-invariance condition for $X$ leads to, 
\begin{equation}
M^{\gamma }\left( X\right) =M^{\gamma }\left( D\left( \mathcal{P}\right)
^{\dagger }XD\left( \mathcal{P}\right) \right) =\mathfrak{D}^{\left( \gamma
\right) }\left( \mathcal{P}\right) ^{\dagger }M^{\gamma }\left( X\right) 
\mathfrak{D}^{\left( \gamma \right) }\left( \mathcal{P}\right) ,
\end{equation}%
which, by the unitarity of $\mathfrak{D}^{\left( \gamma \right) }\left( 
\mathcal{P}\right) $, yields, 
\begin{equation}
\mathfrak{D}^{\left( \gamma \right) }\left( \mathcal{P}\right) M^{\gamma
}\left( X\right) =M^{\gamma }\left( X\right) \mathfrak{D}^{\left( \gamma
\right) }\left( \mathcal{P}\right) ,  \label{X44}
\end{equation}%
for all $\mathcal{P}\in \mathcal{S}_{N}$. For any permutation-invariant
operator $X$ and any $\boldsymbol{a}$ , $\boldsymbol{a}^{\prime }\in
ev\left( \boldsymbol{A},\gamma \right) $, an application of the third
Schur's Lemma to the identity (\ref{X44}) shows that the matrix $M^{\gamma
}\left( X\right) $ is proportional to the identity matrix\footnote{%
The notation, $\left\langle \gamma ,\boldsymbol{a}^{\prime }\left\vert
\left\vert X\right\vert \right\vert \gamma ,\boldsymbol{a}\right\rangle $,
for the proportionality coefficient is chosen by analogy to the reduced
matrix element used in connection with the rotation group.}; therefore,%
\begin{equation}
\left\langle \gamma ,\boldsymbol{a}^{\prime }:g^{\prime }\left\vert
X\right\vert \gamma ,\boldsymbol{a}:g\right\rangle =\left\langle \gamma ,%
\boldsymbol{a}^{\prime }\left\vert \left\vert X\right\vert \right\vert
\gamma ,\boldsymbol{a}\right\rangle \text{ }\delta _{g^{\prime }g}.
\label{RedMe}
\end{equation}%
Applying (\ref{nocon}) and (\ref{RedMe}) to the case $X=1$ yields the
orthonormality conditions, 
\begin{equation}
\left\langle \left. \gamma ^{\prime },\boldsymbol{a}^{\prime }:g^{\prime
}\right\vert \gamma ,\boldsymbol{a}:g\right\rangle =\delta _{\gamma \gamma
^{\prime }}\delta _{\boldsymbol{aa}^{\prime }}\delta _{gg^{\prime }}.
\label{OrthRep}
\end{equation}

The state vectors of symmetry type $\gamma $ form a subspace, $\mathfrak{H}%
_{\gamma }^{\left( N\right) }$, with basis set 
\begin{equation}
\mathfrak{B}\left[ \mathfrak{H}_{\gamma }^{\left( N\right) }\right] =\left\{
\left\vert \gamma ,\boldsymbol{a}:g\right\rangle \ |\ \boldsymbol{a}\in
ev\left( \boldsymbol{A},\gamma \right) ,\ g=1,\ldots ,d_{\gamma }\right\} .
\end{equation}%
As $\gamma $ ranges through $\Gamma _{N}$ the combined basis vectors of the $%
\mathfrak{H}_{\gamma }^{\left( N\right) }$'s provide a basis set for $%
\mathfrak{H}^{\left( N\right) }$. By virtue of (\ref{OrthRep}) the $%
\mathfrak{H}_{\gamma }^{\left( N\right) }$'s are mutually orthogonal and
thus linearly independent. Furthermore, (\ref{nocon}) shows that
permutation-invariant operators cannot connect $\mathfrak{H}_{\gamma
}^{\left( N\right) }$ and $\mathfrak{H}_{\gamma ^{\prime }}^{\left( N\right)
}$ for $\gamma \neq \gamma ^{\prime }$.These properties combine to yield the,

\begin{quote}
\textbf{Superselection Rule}: For any permutation-invariant operator $A$, 
\begin{equation}
\ A:\mathfrak{H}_{\gamma }^{\left( N\right) }\rightarrow \mathfrak{H}%
_{\gamma }^{\left( N\right) }\text{ and }\mathfrak{H}^{\left( N\right)
}=\dbigoplus\limits_{\gamma \in \Gamma _{N}}\mathfrak{H}_{\gamma }^{\left(
N\right) }\   \label{DsumSS}
\end{equation}
\end{quote}

\noindent The subspaces $\mathfrak{H}_{\gamma }^{\left( N\right) }$ are
called superselection sectors\ \cite[Chap III.1]{RH}. Since observables and
physical density operators are permutation-invariant, they do not connect
different superselection sectors.

The superselection rule has significant consequences for time evolution. The
unitary time evolution operator, $U\left( t\right) :=\exp \left(
-itH/\hslash \right) $, is permutation-invariant by virtue of the
permutation-invariance of the Hamiltonian; therefore, it satisfies (\ref%
{nocon}). Thus $\left\langle \Phi \left\vert U\left( t\right) \right\vert
\Psi \right\rangle =0$ if $\left\vert \Phi \right\rangle $\ and $\left\vert
\Psi \right\rangle $\ belong to different superselection sectors. In
particular, this means that $U\left( t\right) \left\vert \Psi ^{\gamma
}\left( 0\right) \right\rangle $ remains in $\mathfrak{H}_{\gamma }^{\left(
N\right) }$ at all later times, 
\begin{equation}
\left\vert \Psi ^{\gamma }\left( t\right) \right\rangle :=U\left( t\right)
\left\vert \Psi ^{\gamma }\left( 0\right) \right\rangle \in \mathfrak{H}%
_{\gamma }^{\left( N\right) }\text{ for all }t>0,  \label{evol99}
\end{equation}%
so that the terms of different symmetry type in (\ref{expgam88}) evolve
independently, 
\begin{equation}
\left\vert \Psi \left( t\right) \right\rangle =U\left( t\right) \left\vert
\Psi \left( 0\right) \right\rangle =\sum_{\gamma }C_{\gamma }U\left(
t\right) \left\vert \Psi ^{\gamma }\left( 0\right) \right\rangle
=\sum_{\gamma }C_{\gamma }\left\vert \Psi ^{\gamma }\left( t\right)
\right\rangle .  \label{Psi(t)}
\end{equation}

Since SP-violations are expected to be small, it is safe to assume that
there is a dominant term, $C_{\gamma _{FB}}$, in (\ref{expgam88}) and (\ref%
{Psi(t)}), where either $\gamma _{FB}=F$ (Fermi representation) or $\gamma
_{FB}=B$ (Bose representation). Thus the strength of possible SP-violations
is determined by the initial amplitudes, $C_{\gamma }$, for $\gamma \neq
\gamma _{FB}$. Note that combining (\ref{PSauto})--which shows that every
operator in the SPQM version satisfies the necessary condition (\ref%
{PermInvOp}) for observables--with (\ref{evol99})--which shows that time
evolution preserves symmetry type--implies that the SPQM version can be
completely recovered from the IPQM version by simply setting $C_{\gamma }=0$
for all $\gamma \neq \gamma _{FB}$. In other words, the Symmetrization
Postulate is equivalent to the equally mysterious condition that the only
superselection sector present in any initial state is either $\gamma =F$ or $%
\gamma =B$. A less stringent assumption is that $\left\vert C_{\gamma
}\right\vert <<\left\vert C_{\gamma _{FB}}\right\vert \ $for all $\gamma
\neq \gamma _{FB}$. Combining this with the normalization condition (\ref%
{norm33}) yields 
\begin{equation}
\left\vert C_{\gamma }\right\vert <<1\ \text{for all }\gamma \neq \gamma
_{FB},  \label{WeakOther}
\end{equation}%
which again says that violations of the Symmetrization Postulate are rare.

The superselection rules commonly encountered in quantum theory are\
associated with continuous symmetries, e.g. rotations and gauge
transformations. Both of these examples are often said to impose
restrictions on the superposition principle. For example, superpositions of
states with integer and half-integer total angular momentum, or
superpositions of states with different net charges are both said to be
forbidden. As explained in \cite[Chap III.1]{RH} another way to understand
this situation is that these superpositions are not forbidden; rather, they
act as mixed states for all observables. The superselection rule (\ref%
{DsumSS}) has the same effect for the discrete symmetry group $\mathcal{S}%
_{N}$. The expectation value of any permutation-invariant operator $X$ for a
general pure state $\left\vert \Psi \right\rangle $ is,%
\begin{equation}
\left\langle \Psi \left\vert X\right\vert \Psi \right\rangle =\sum_{\gamma
^{\prime }}\sum_{\gamma }C_{\gamma ^{\prime }}^{\ast }C_{\gamma
}\left\langle \Psi ^{\gamma ^{\prime }}\left\vert X\right\vert \Psi ^{\gamma
}\right\rangle =\sum_{\gamma }\left\vert C_{\gamma }\right\vert
^{2}\left\langle \Psi ^{\gamma }\left\vert X\right\vert \Psi ^{\gamma
}\right\rangle ,  \label{ExpV}
\end{equation}%
where the final form is a consequence of the superselection rule. Since (\ref%
{ExpV}) holds for all observables and only involves the magnitudes $%
\left\vert C_{\gamma }\right\vert $, there are no interference terms between
states of different symmetry type and no information about the phases of the
coefficients, $C_{\gamma }$, can be obtained from any measurement of
observables, Even though $\left\vert \Psi \right\rangle $ is a
hybrid-symmetry pure state, (\ref{ExpV}) shows that the expectation value of
any observable is an average over a statistical mixture of the
definite-symmetry pure states included in $\left\vert \Psi \right\rangle $.
In other words, for evaluating averages of observables the pure state $%
\left\vert \Psi \right\rangle $ acts as a mixed state described by the
permutation-invariant density operator,%
\begin{equation}
\rho _{\Psi }:=\sum_{\gamma }\left\vert \Psi ^{\gamma }\right\rangle
\left\vert C_{\gamma }\right\vert ^{2}\left\langle \Psi ^{\gamma
}\right\vert ,  \label{HsymPureMixed}
\end{equation}

\section{Search for SP-violations\label{&Experiment}}

The strong conditions imposed on state vectors by the Symmetrization
Postulate form the basis of the conventional (SPQM) description of all
systems of identical particles, ranging from a small number of particles
involved in a scattering event to interacting many-body systems, e.g.
Bose-Einstein condensates, superfluids, superconductors, etc.. This broad
range of influence implies an equally broad range of possible experiments to
search for SP-violations. The formalism developed in the previous sections
could, for example, be used to construct an extended version of quantum
statistical mechanics that is not limited to Bose or Fermi statistics.
However, experiments on many-body systems may not be the most useful
approach. The difficulty is that the small size of SP-violations would very
likely produce subtle effects that would be extremely difficult to detect.
This may be the reason that experiments for both Fermi systems \cite%
{FSP1,FSP2,FSP3,FSP4,FSP5} and Bose systems \cite{FSP6,FSP7} typically
involve interactions of single electrons or photons with atoms or molecules.
While detection of SP-violating events for these systems will also be
extremely difficult, these experiments can take advantage of selection rules
imposed by Bose or Fermi statistics. For example, in SPQM the initial and
final states for an electron (photon) scattering event cannot include a
symmetric (antisymmetric) state of two electrons (photons)

In order to obtain an observable signal from a weak violation of an
SP-imposed selection rule, it is essential to have a large flux of incident
particles. One way to achieve this with electrons is to induce a strong flow
of current through a conductor. Experiments using this arrangement to test
for PEP-violations have been conducted and carefully analyzed \cite{FSP4}.
The idea of these experiments is that a radiation cascade would occur when a
conduction electron is captured by an atom in the crystal lattice. Captures
producing X-rays would, however, be PEP-forbidden since the relevant lower
energy levels are fully occupied. Thus an emission of X-rays in the
appropriate energy range would be a signal of a PEP-violation. A model for
this event must evidently go beyond SPQM, and the IPQM version of quantum
mechanics provides a minimal extension suited to this purpose.

In order to illustrate how the unfamiliar aspects of IPQM are involved in an
analysis of such experiments, it is useful to consider a simplified model.
The relevant issues can not arise for single particles and are essentially
trivial for two-particle systems, which only support the symmetric and
antisymmetric representations. Thus the simplest model systems that are
informative are those with three indistinguishable particles. These
considerations suggest the following toy model. The incident electron and
the electrons in the target atom are modeled by three identical,
non-interacting, spin-1/2 particles confined to one space dimension.
Unperturbed particle dynamics are described by a spin-independent
single-particle Hamiltonian $H_{0}$, and the coupling to the radiation field
is given by $H_{rad}$. The single-particle quantum numbers are $\theta
=\left( \epsilon ,s\right) $, where $\epsilon $ and $s\hslash $ are
respectively eigenvalues of $H_{0}$ and the $z$-component of the spin. In
the initial state, the atom is modeled by two particles in the ground state
of $H_{0}$, with total energy $2\epsilon _{0}$, and the incident electron is
modeled by one particle in an excited state, with energy $\epsilon
_{1}>\epsilon _{0}$. The final state has all three particles in the ground
state. For distinguishable particles this situation could be described by
assigning $\left\vert \boldsymbol{\theta }_{int}\right\rangle =\left\vert
\left( \epsilon _{0},1/2\right) ,\left( \epsilon _{0},-1/2\right) ,\left(
\epsilon _{1},1/2\right) \right\rangle $, and $\left\vert \boldsymbol{\theta 
}_{fin}\right\rangle =\left\vert \left( \epsilon _{0},1/2\right) ,\left(
\epsilon _{0},-1/2\right) ,\left( \epsilon _{0},1/2\right) \right\rangle $
as nominal initial and final state vectors respectively.

In the SPQM\ version for fermions, the state vectors $\left\vert \boldsymbol{%
\theta }_{int}\right\rangle $ and $\left\vert \boldsymbol{\theta }%
_{fin}\right\rangle $ would be replaced by the antisymmetrized states,%
\begin{eqnarray}
\left\vert F,\boldsymbol{\theta }_{int}\right\rangle &:&=\frac{1}{\sqrt{3!}}%
\sum_{\mathcal{P}\in \mathcal{S}_{3}}s_{\mathcal{P}}D\left( \mathcal{P}%
\right) \left\vert \boldsymbol{\theta }_{int}\right\rangle ,  \notag \\
\left\vert F,\boldsymbol{\theta }_{fin}\right\rangle &:&=\frac{1}{\sqrt{3!}}%
\sum_{\mathcal{P}\in \mathcal{S}_{3}}s_{\mathcal{P}}D\left( \mathcal{P}%
\right) \left\vert \boldsymbol{\theta }_{fin}\right\rangle =0,  \label{qpf=0}
\end{eqnarray}%
where $s_{\mathcal{P}}=\left( +1,-1\right) $ for (even, odd) $\mathcal{P}$.
The second equation is an example of the rule that two identical fermions
cannot occupy the same single-particle state.

In the IPQM\ version all three irreducible representations of $\mathcal{S}%
_{3}$, \emph{cf.} Appendix \ref{&S3}, must be considered: $\gamma =F$ , $%
\gamma =B$ , and $\gamma =I$. The irreducible representations that overlap
an unsymmetrized three-particle state $\left\vert \boldsymbol{\kappa }%
\right\rangle $ are determined by calculating the normalized projection of $%
\left\vert \boldsymbol{\kappa }\right\rangle $ onto a basis vector in a
carrier space for the irreducible representation $\mathfrak{D}^{\left(
\gamma \right) }$, i.e., 
\begin{equation}
\left\vert \gamma ;\boldsymbol{\kappa }:g\right\rangle :=\frac{\Pi _{\gamma
g}\left\vert \boldsymbol{\kappa }\right\rangle }{\sqrt{\left\langle 
\boldsymbol{\kappa }\left\vert \Pi _{\gamma g}\right\vert \boldsymbol{\kappa 
}\right\rangle }},  \label{irepgg}
\end{equation}%
where $\Pi _{\gamma g}$ is one of the projection operators, $\Pi _{B}$,$\
\Pi _{F}$, and $\Pi _{Ig}$, defined in (\ref{PiB})-(\ref{PiI}). Evaluating (%
\ref{irepgg}) for $\left\vert \boldsymbol{\theta }_{int}\right\rangle $
yields, 
\begin{eqnarray}
\left\vert B,\boldsymbol{\theta }_{int}:1\right\rangle &\propto &\Pi
_{B}\left\vert \boldsymbol{\theta }_{int}\right\rangle \neq 0,  \notag \\
\left\vert F,\boldsymbol{\theta }_{int}:1\right\rangle &\propto &\Pi
_{F}\left\vert \boldsymbol{\theta }_{int}\right\rangle \neq 0,  \notag \\
\left\vert I,\boldsymbol{\theta }_{int}:g\right\rangle &\propto &\Pi
_{Ig}\left\vert \boldsymbol{\theta }_{int}\right\rangle \neq 0,
\end{eqnarray}%
which means that all three representations overlap with $\left\vert 
\boldsymbol{\theta }_{int}\right\rangle $. For $\left\vert \boldsymbol{%
\theta }_{fin}\right\rangle $, 
\begin{eqnarray}
\left\vert B,\boldsymbol{\theta }_{fin}:1\right\rangle &\propto &\Pi
_{B}\left\vert \boldsymbol{\theta }_{fin}\right\rangle \neq 0,  \notag \\
\left\vert I,\boldsymbol{\theta }_{fin}:g\right\rangle &\propto &\Pi
_{Ig}\left\vert \boldsymbol{\theta }_{fin}\right\rangle \neq 0,  \notag \\
\left\vert F,\boldsymbol{\theta }_{fin}:1\right\rangle &\propto &\Pi
_{F}\left\vert \boldsymbol{\theta }_{fin}\right\rangle =0,
\end{eqnarray}%
i.e. only the representations $B$ and $I$ overlap with $\left\vert 
\boldsymbol{\theta }_{fin}\right\rangle $. The general normalized states
that can describe this experiment are an initial state%
\begin{equation}
\left\vert \Psi _{int}\right\rangle =C_{int}^{F}\left\vert F,\boldsymbol{%
\theta }_{int}:1\right\rangle +C_{int}^{B}\left\vert B,\boldsymbol{\theta }%
_{int}:1\right\rangle +\sum_{g=1}^{2}C_{int}^{Ig}\left\vert I,\boldsymbol{%
\theta }_{int}:g\right\rangle ,  \label{instate}
\end{equation}%
and a final state,%
\begin{equation}
\left\vert \Psi _{fin}\right\rangle =C_{fin}^{B}\left\vert B,\boldsymbol{%
\theta }_{fin}:1\right\rangle +\sum_{g=1}^{2}C_{fin}^{Ig}\left\vert I,%
\boldsymbol{\theta }_{fin}:g\right\rangle .  \label{outstate}
\end{equation}

A detectable PEP-violating event for this model would be the emission of
radiation at the resonance frequency $\omega $ given by $\hslash \omega
=\left( 2\epsilon _{0}+\epsilon _{1}\right) -\left( 3\epsilon _{0}\right)
=\epsilon _{1}-\epsilon _{0}$. Since $H_{rad}$ is permutation invariant, the
general result (\ref{RedMe}) and the superselection rule (\ref{DsumSS})
combine to yield the transition matrix element,%
\begin{eqnarray}
\left\langle \Psi _{fin}\left\vert H_{rad}\right\vert \Psi
_{int}\right\rangle &=&C_{fin}^{B\ast }C_{int}^{B}\ \left\langle B,%
\boldsymbol{\theta }_{fin}\left\vert H_{rad}\right\vert B,\boldsymbol{\theta 
}_{int}\right\rangle  \notag \\
&&+\left\{ \sum_{g=1}^{2}C_{fin}^{Ig\ast }\ C_{int}^{Ig}\right\} \
\left\langle I,\boldsymbol{\theta }_{fin}\left\vert \left\vert
H_{rad}\right\vert \right\vert I,\boldsymbol{\theta }_{int}\right\rangle .
\label{MEbad}
\end{eqnarray}%
which shows that the probability of detecting a PEP-violation is determined
by the initial amplitudes $C_{int}^{B}$ and $C_{int}^{Ig}$ for the Bose and
mixed representations of $\mathcal{S}_{3}$. This is an explicit example of
the general conclusion following from (\ref{evol99}). When the initial state
vector satisfies (\ref{WeakOther}), the expansion coefficients satisfy $%
\left\vert C_{int}^{B}\right\vert <<1$ and $\left\vert
C_{int}^{Ig}\right\vert <<1$, so that the transition rate is small.

\section{Discussion\label{&IPQM}}

The IPQM\ version of quantum mechanics is complicated by the necessity of
considering all irreducible representations of $\mathcal{S}_{N}$. As shown
in Section \ref{&Filtering} this requires changes in the usual description
of state preparation. The most significant new feature of IPQM\ is the
presence of the superselection rule associated with the discrete symmetry
group $\mathcal{S}_{N}$. As shown in Section \ref{&SSR} this has important
effects on time evolution and the interpretation of superpositions of states
of different symmetry type. These features of IPQM are consequences of three
assumptions: (A1) The axioms of quantum theory. (A2) The definition of
identical particles. (A3) The Indistinguishability Postulate. These
assumptions lead to the following properties: (1) Permutation symmetries
hold without regard to interactions. (2) Permutation symmetries hold for all
interparticle separations. (3) The symmetry type of a state is preserved by
time evolution. The first two properties are purely kinematical, and the
third follows from (\ref{evol99}).

The analysis of the toy model of a PEP-violating experiment in Section \ref%
{&Experiment} is focussed on the implications of properties (1)-(3) of IPQM,
but an alternative description has been proposed for this class of
experiments. In this approach the target and incident electrons are treated
as separate systems, and the incident electrons are said to be \emph{fresh} 
\cite{FSP4}. The meaning of\ the term `fresh electrons' is implicit in the
following assumptions: (i) The fresh electrons have never interacted with
the target electrons. (ii) No symmetry conditions have been established
between the fresh electrons and the target electrons. (iii) Establishment of
any symmetry conditions between the fresh electrons and\emph{\ }the target
electrons depends on interactions between them. The assumptions (i)-(iii) of
the fresh-electron model are clearly inconsistent with properties (1)-(3) of
IPQM and thus with one or all of the assumptions (A1)-(A3). This does not
mean that the fresh electron assumptions are unphysical, but it does imply
that they require an extension of SPQM that is substantially more radical
than IPQM. This is not necessarily a bad feature; changing something as
fundamental as the Pauli Exclusion Principle may well require a radically
new theory.

The decision to formulate IPQM using nonrelativistic quantum theory was not
made for the sake of simplicity; indeed, it cannot be avoided. It is not
possible to formulate a relativistic quantum theory with a fixed number of
particles \cite[Chap 1.1]{SW}; consequently, any theory that assumes a fixed
number of particles--in particular IPQM--is necessarily nonrelativistic. The
fact that IPQM does not impose any connection between spin and statistics 
\cite{MG}, may also be related to the use of nonrelativistic quantum theory.
The spin-statistics connection has only been established by using an
argument based on relativistic quantum field theory \cite{RH,CGH} to prove
the \textbf{spin-statistics theorem}:\textbf{\ }\emph{integer spin particles
are bosons and half-integer spin particles are fermions}.\ The statement of
this theorem seems to suggest that only Bose and Fermi statistics are
possible in a relativistic theory, but this is misleading. The various
proofs of the spin-statistics theorem depend on the following assumption: : 
\emph{quantum fields evaluated at spacelike separated points either commute
or anti-commute} (CorAC). In some proofs this assumption is included in the
hypothesis of the theorem and in others it is found in the associated axioms
of field theory. In either case CorAC is used to impose the empirical
restriction to Bose or Fermi statistics on relativistic quantum field
theory, just as the Symmetrization Postulate imposes it on nonrelativistic
quantum mechanics. If it could be demonstrated that CorAC follows from the
axioms of any consistent relativistic quantum field theory, then the
conclusion would be that only Bose or Fermi statistics are possible. If,
instead, there is a consistent field theory for which CorAC could be false,
then it might be possible to introduce other kinds of statistics into field
theory.

\appendix

\section{The Symmetric Group\label{&IRepSN}}

The Symmetric Group $\mathcal{S}_{N}$ is of order $N!$ and each permutation
in $\mathcal{S}_{N}$ can be written as the product of transpositions. A
permutation is even (odd) if it is the product of an even (odd) number of
transpositions. The parity of a permutation $\mathcal{P}$ is $s_{\mathcal{P}%
}=\left( +1,-1\right) $ for (even, odd) $\mathcal{P}$.

Each irreducible representation , $\mathcal{P}\rightarrow \mathfrak{D}%
^{\left( \gamma \right) }\left( \mathcal{P}\right) $, of $\mathcal{S}_{N}$
is finite dimensional and the labels, $\gamma :=\left( \lambda _{1},\ldots
,\lambda _{N}\right) $, range over the finite set, $\Gamma _{N}$, of
solutions of $\lambda _{1}+\cdots +\lambda _{N}=N$, subject to $\lambda
_{n}\geq \lambda _{n+1}\geq 0$. For each $N$ there are only two
one-dimensional representations: The symmetric Bose representation, $\gamma
=B=\left( N,0,\ldots ,0\right) $, and the antisymmetric Fermi
representation, $\gamma =F=\left( 1,1,\ldots ,1\right) $. Projection
operators, $\Pi _{\gamma g}$,\ satisfying, 
\begin{equation}
\Pi _{\gamma ^{\prime }g^{\prime }}\Pi _{\gamma g}=\delta _{\gamma ^{\prime
}\gamma }\delta _{g^{\prime }g}\Pi _{\gamma g}\ \ \left( \gamma ,\gamma
^{\prime }\in \Gamma _{N},\ g=1,\ldots ,d_{\gamma }\right) ,
\end{equation}%
are used to define basis vectors for a space $\mathfrak{N}$ carrying an
irreducible representation of $\mathcal{S}_{N}$ \cite[Chap. 7]{MH}.

\subsection{Schur's Lemmas.\label{&SchurL}}

The following statements hold for any finite group $G$ \cite[Chap 3.14]{MH}.

\begin{lemma}
\label{!SL1}Let $\mathfrak{D}\left( g\right) $ and $\mathfrak{D}^{\prime
}\left( g\right) $ be matrices of the irreducible representations $\mathfrak{%
D}$ and $\mathfrak{D}^{\prime }$ of $G$ with different dimensions, then a
matrix $M$ satisfying $\mathfrak{D}\left( g\right) M=M\mathfrak{D}^{\prime
}\left( g\right) $ for all $g\in G$ \ necessarily vanishes.
\end{lemma}

\begin{lemma}
\label{!SL2}Let $\mathfrak{D}\left( g\right) $ and $\mathfrak{D}^{\prime
}\left( g\right) $ be matrices of the irreducible representations $\mathfrak{%
D}$ and $\mathfrak{D}^{\prime }$ with the same dimension. If a matrix $M$
satisfies $\mathfrak{D}\left( g\right) M=M\mathfrak{D}^{\prime }\left(
g\right) $ for all $g\in G$, then either $\mathfrak{D}\left( g\right) $ and $%
\mathfrak{D}^{\prime }\left( g\right) $\ are equivalent or $M=0$.
\end{lemma}

\begin{lemma}
\label{!SL3}If the matrices $\mathfrak{D}\left( g\right) $ provide an
irreducible representation of $G$ and $M\mathfrak{D}\left( g\right) =%
\mathfrak{D}\left( g\right) M$ for all $g\in G$, then $M$ is a multiple of
the identity matrix.
\end{lemma}

\subsection{Symmetric Group\ of order 3\label{&S3}}

\textbf{Group elements}: The $3!$ permutations in $\mathcal{S}_{3}$ are $%
\left\{ e,\mathcal{P}_{12},\mathcal{P}_{13},\mathcal{P}_{23},\mathcal{P}%
_{132},\mathcal{P}_{123}\right\} $, where,

$e\left( n\right) =n$; $\mathcal{P}_{ij}\left( j\right) =i$; $\mathcal{P}%
_{ij}\left( i\right) =j$; $\mathcal{P}_{ij}\left( n\right) =n$ for $n\neq
i,j $ ;

$\mathcal{P}_{132}\left( 1\right) =3$, $\mathcal{P}_{132}\left( 3\right) =2$%
, $\mathcal{P}_{132}\left( 2\right) =1$;

$\mathcal{P}_{123}\left( 1\right) =2$, $\mathcal{P}_{123}\left( 2\right) =3$%
, $\mathcal{P}_{123}\left( 3\right) =1$.

\noindent \textbf{Irreducible representations}

The labels $\gamma =\left( \lambda _{1},\lambda _{2},\lambda _{3}\right) $
in $\Gamma _{3}$ satisfy $\lambda _{1}+\lambda _{2}+\lambda _{3}=3$ with $%
\lambda _{1}\geq \lambda _{2}\geq \lambda _{3}\geq 0.$ There are three
solutions: $\gamma =\left( 3,0,0\right) =B$, $\gamma =\left( 1,1,1\right) =F$%
, and $\gamma =\left( 2,1,0\right) =I$. The first two respectively label the
one-dimensional Bose and Fermi representations, and the third labels a
two-dimensional representation of intermediate symmetry. The projection
operators\ are given by, 
\begin{equation}
\text{Bose}\text{: }\Pi _{\left( 3,0,0\right) }=\Pi _{B}:\mathfrak{=}\frac{1%
}{3!}\sum_{\mathcal{P}\in \mathcal{S}_{3}}D\left( \mathcal{P}\right) ,
\label{PiB}
\end{equation}%
\begin{equation}
\text{Fermi}\text{: }\Pi _{\left( 1,1,1\right) }=\Pi _{F}\ :=\frac{1}{3!}%
\sum_{\mathcal{P}\in \mathcal{S}_{3}}s_{\mathcal{P}}D\left( \mathcal{P}%
\right) ,  \label{PiF}
\end{equation}%
\begin{eqnarray}
\text{Intermediate} &\text{:}&\Pi _{Ig}=\Pi _{\left( 2,1,0\right) g}\ \left(
g=1,2\right) ,  \notag \\
\Pi _{I1} &:&=\frac{1}{3}\left( \hat{e}+D\left( \mathcal{P}_{12}\right) 
\mathcal{-}D\left( \mathcal{P}_{13}\right) \mathcal{-}D\left( \mathcal{P}%
_{123}\right) \right) ,  \notag \\
\ \Pi _{I2} &:&=\frac{1}{3}\left( \hat{e}-D\left( \mathcal{P}_{12}\right)
+D\left( \mathcal{P}_{13}\right) -D\left( \mathcal{P}_{132}\right) \right) .
\label{PiI}
\end{eqnarray}

\begin{acknowledgement}
I would like to thank Professor Dmitry Budker, whose wide ranging interests
in fundamental issues of physics rekindled my own interest in the subject of
particle statistics.
\end{acknowledgement}

\end{document}